# μSR Study of Spin Dynamics in $LiY_{1-x}Ho_xF_4$

R. C. Johnson[1], K. Chen[1], S. R. Giblin[2], J. S. Lord[2], A. Amato[3], C. Baines[3],
B. Barbara[4], B. Z. Malkin[5], and M. J. Graf[1]

[1] Department of Physics, Boston College, Chestnut Hill, MA 02467 USA
[2] Rutherford Appleton Laboratory, Didcot, Oxfordshire OX11 0QX, UK
[3] Paul Scherrer Institute, CH 5232 Villigen PSI, Switzerland
[4] Néel Institute, Dept. of Nanosciences, CNRS, 38042 Grenoble Cedex-09, France
[5] Kazan Federal University, Kazan 420008, Russian Federation

We present zero-field μSR measurements for $LiY_{1-x}Ho_xF_4$ samples with $x$ = 0.0017, 0.0085, 0.0406, and 0.0855. We characterize the dynamics associated with the formation of the $(F-\mu-F)^{-1}$ complex by comparing our data to Monte Carlo simulations to determine the concentration range over which the spin dynamics are determined primarily by the $Ho^{3+}$-μ interaction rather than the F-μ interaction. Simulations show that F-μ-F oscillations should evolve into a Lorentzian Kubo-Toyabe decay for an increasing static magnetic field distribution $\Gamma$ (*i.e.*, increasing $x$), but the data do not show this behavior, consistent with the recently reported existence of strong magnetic fluctuations in this system at low temperatures. Anisotropy in the field distribution is shown to cause small errors of order 10% from behavior predicted for an isotropic distribution. Finally, numerical calculations show that values of $\Gamma$ calculated in the single ion limit greatly exceed the values extracted from curve fits, suggesting that strong correlations play an important role in this system.

Magnetism in $LiY_{1-x}Ho_xF_4$ has been studied for decades. Pure $LiHoF_4$ is an Ising ferromagnet, with Curie temperature $T_C$ = 1.53 K [1]. Dilution of Ho by non-magnetic Y frustrates magnetic interactions, suppressing $T_C$. Near $x$ = 0.25 a low-temperature spin glass state has been reported [2], and an unusual 'anti-glass' state was reported for $x$ = 0.045 [3]. In the very low $x$ regime, the system is a single-ion magnet, with low temperature spin dynamics dominated by field-induced avoided level crossings and exhibiting quantum tunneling of the magnetization [4]. Much of the low temperature physics is understandable through the relatively simple energy diagram for the $Ho^{3+}$ ion, where the lowest multiplet $^5I_8$ is split by the crystal electric field into a ground state Ising doublet separated from the first excited singlet state by a gap of 10 K; hyperfine coupling with the $I$ = 7/2 $^{165}$Ho (100% abundance) nucleus produces further structure in the ground state manifold and results in magnetic field induced avoided level crossings.

The samples are large, high quality single crystals, with well-studied physical parameters, making the system amenable to numerical simulations [5, 6] to determine the relative importance of factors controlling the spin dynamics, such as spin-phonon coupling and Ho-Ho correlations. μSR is a powerful probe of spin dynamics and has been successfully used to study the single-ion dynamics [7] for very dilute samples ($x$ = 0.002). Recently μSR was used to study the possible onset of spin glass ordering [8] for samples with concentrations in the range $0.018 \leq x \leq 0.25$, with the authors concluding that there was no evidence for spin glass transition, and that strong magnetic fluctuations are present at low temperatures for all concentrations.

Here we extend our μSR earlier measurements by increasing the Ho concentration in order to study the onset of correlated magnetism in this system in a 'bottom up' approach. These results complement those of Ref. 8, and confirm their conclusion that fluctuations remain strong in this system even at low temperatures. Additionally, we characterize the dynamics associated with the formation of the bonded $(F-\mu-F)^{-1}$ ion ("FμF"), by comparing our results to numerical simulations, allowing us to determine the concentration and applied magnetic field range for which the spin dynamics are determined primarily by the Ho-μ interaction rather than the F-μ interaction. Finally we show that detailed numerical simulations for the time dependent muon response are required to extract quantitative information regarding the Ho spin dynamics.

We studied large single crystal samples with concentrations $x$ = 0.0017, 0.0085, 0.0406, and 0.0855 as determined by DC magnetization measurements at 2 K. The magnetic field variation of the AC susceptibility at 2 K was also studied, and the $x$ = 0.0017 sample exhibited sharp features at (and midway between) the avoided level crossings, agreeing fully with previous results [4, 5].

In time-differential μSR, spin polarized muons are implanted in the sample, and the time dependence of the muon spin orientation can be monitored via the positrons emitted from millions of muon decay events, from which one extracts the asymmetry $A(t)$ [9]. The asymmetry is proportional to the muon spin autocorrelation function (or depolarization function) $G(t)$. The time dependence of $G$ results from quasi-static processes, for example the spatial disorder present in the local magnetic field, and by dynamical processes, such as local field fluctuations. Experiments were run at the pulsed muon source ISIS and the continuous muon source at the Paul Scherrer Institute. The ISIS data were taken in conventional polarization mode, where the muons are 100% polarized with the spin directed antiparallel to the beam momentum. The PSI data were taken in spin rotated mode, where the muon spin is oriented at approximately 50° to the beam momentum, allowing for simultaneous measurement of the depolarization for spin components in directions parallel and perpendicular to the beam momentum. Data at both facilities were taken over the temperature range 1.8 K < $T$ < 50 K in gas flow cryostats, with a limited amount of data taken in dilution refrigerators.

For all samples the zero field depolarization at 50 K shows the strong features characteristic of weakly damped coherent spin oscillations of the linear FμF [10] (see Fig. 1). Neglecting F-F coupling, there are oscillatory components at three frequencies (excluding the zero-frequency mode). The relative amplitudes of the components will depend on the angle $\theta$ between the initial muon spin direction and the axis of the FμF complex, and the depolarization function is given by

$$G_{F\mu F}(t) = \frac{1}{12}\{5+3\zeta + (1+3\zeta)\cos(\sqrt{3}\omega_d t) + 2(1-\zeta)[\frac{\omega_2}{\omega_d}\cos(\omega_2 t) + \frac{\omega_3}{\omega_d}\cos(\omega_3 t)]\}, \quad (1)$$

where the characteristic frequency $\omega_D$ is given by $\omega_D = \frac{\hbar\gamma_\mu\gamma_F}{r^3}$, $\omega_2 = (3-\sqrt{3})\omega_d/2$, $\omega_3 = (3+\sqrt{3})\omega_d/2$, $\gamma_\mu$ and $\gamma_F$ are the gyromagnetic ratios for the muon and fluorine nucleus, respectively, $\zeta = \cos^2\theta$, and $r$ is the distance between a muon and the nearest neighbor fluorine nucleus. Assuming light damping, the high temperature asymmetry data were fit to the form

$$A(t) = A_1 G_{F\mu F} \exp\left[-(\lambda_{ZF} t)^\beta\right] + A_2 \exp(-\lambda_2 t)^2 + A_{BG} , \qquad (2)$$

and we find for $x = 0.0017$ that $\lambda_{ZF} = 0.17 \pm 0.01$ µs$^{-1}$, $\beta = 1.16 \pm 0.04$, $\omega_D = 1.31 \pm 0.01$ MHz, and $\theta = 62 \pm 5$ ° (48 ± 4 °) for the muon polarization parallel (perpendicular) to the crystalline c-axis; the overall $\chi^2$ per degree of freedom was typically 1.2. The term $A_2$ accounts for muons not forming FµF in the sample (roughly 15% of the total relaxing signal), and $A_{BG}$ accounts for muons which are imbedded in the silver sample holders at ISIS (5-10% of the total asymmetry). It is difficult to interpret the value of $\lambda_2 = 0.48 \pm 0.014$ µs$^{-1}$ due to the small size of $A_2$, which results in an insensitivity to the exact form for the relaxation function used. For example, setting the exponent in the second term to one only slightly increases $\chi^2$ to 1.25; the amplitudes and other fit parameters remain unchanged, apart from the exponent $\beta$ in the FµF term which then increases to 1.4. The value of $\theta = 62°$ is close to the angle $\theta_0 = 68.8°$ between the muon spin and FµF line for the most likely muon stopping sites, the four interstitial locations per unit cell (0 ±1/4 1/8), (±1/4 0 -1/8) located between the pairs of fluorine nuclei with the closest spacing. The value of $\theta = 48°$ coincides with the angle 48.75° corresponding to the mean value of $\zeta$ for these sites with the muon polarization perpendicular to the crystalline c-axis. These angles are approximate, as the FµF line is not along a symmetry axis and some twisting may occur. The mean separation of the fluorine ions extracted from $\omega_D$ is 0.237 nm, representing a local contraction of the fluorine pairs by 8% from their unperturbed separation [11]. Nearly identical results are obtained for all concentrations, showing that the fast fluctuations of the Ho at high temperature are outside the muon frequency domain.

In Fig. 1a we show the temperature evolution of the $x = 0.0017$ depolarization curves along with the fits to Eq. 2; in Fig. 1b we show comparable data for the $x = 0.0854$ sample; the solid lines are for Eq. 2 (T = 50 K, 20 K and 14 K) and a stretched exponential function (Eq. 2, with $G_{F\mu F}(t) = 1$) for lower temperatures. Figure 2 shows the temperature dependence of the fit parameters $\lambda_{ZF}$, $\omega_D$, and $\beta$ (inset) for the $x = 0.0017$ sample in the parallel configuration. The curves were fit with $\theta$ and the A-coefficients fixed at their high temperature values. The parameter values are essentially identical for the muon spin parallel and perpendicular to the crystalline c-axis. $\lambda_{ZF}$ increases rapidly below roughly 10 K, followed by a gradual increase below 2 K. This overall behavior indicates a crossover from depolarization by randomly oriented quasi-static nuclear moments at high temperatures towards a dilute distribution of quasi-static Ho moments as thermally excited spin fluctuations over the electronic energy barrier ~ 10 K and hyperfine splittings (~ 0.2 – 1.4 K) are frozen out. The variation of $\beta$ with temperature is also consistent with this model: at intermediate temperatures $\beta < 1$ indicating strong fluctuations [12], while at low temperatures $\beta \sim 1$. The sharp increase below 6 K in $\omega_D$ is striking, and we believe the apparent increase results from the enhanced quasi-static disorder. As a test, we ran Monte Carlo simulations of the depolarization for a single FµF with parameters similar to ours using the Quantum simulation program [13] and assuming a static isotropic Lorentzian distribution of fields of width $\Gamma$. Some of the depolarization curves for increasing $\Gamma$ (1 to 80 G) are shown in Fig. 3. Fitting the simulated data to Eq. 2 (with no background contribution) over the small $\Gamma$ range 0 to 4 G shows that extracted parameter $\omega_D$ does in fact increase with increasing $\Gamma$.

Overall, from Fig. 3 we see an evolution from nearly undamped FµF oscillations to the (quasi-static) Lorentzian Kubo-Toyabe (LKT) decay [12]

$$G_{KL}(t) = \frac{1}{3} + \frac{2}{3}(1-at)\exp(-at). \qquad (3)$$

A comparison with Fig. 1a shows that the $x = 0.0017$ data at 1.8 K can be roughly characterized by an 8 G Lorentzian distribution, corresponding to $a = 0.7$ µs$^{-1}$ ($a = \gamma_\mu \Gamma$). The undamped FµF depolarization (Eq. 1) has evolved to LKT depolarization (Eq. 2) for $\Gamma$ well above 50 G ($a \approx 4.1$ µs$^{-1}$); neither Eq. 2 nor Eq. 3 fit the simulated data in the range 5 G $< \Gamma <$ 50 G. Fitting the 50 G simulation to Eq. 3 yields errors for the characteristic minimum $t_{min} = 2/a$ and the long-time asymptotic value of 1/3 of about 10%. While the simulations in the low and high damping limits can be adequately described by damped FµF and LKT depolarization functions, respectively, the intermediate regime has no obvious analytical form.

Figure 4 shows the experimental depolarization curves, with muon polarization parallel to the c-axis, at $T = 80$ mK. The increased damping with increasing holmium concentration $x$ is evident. The solid lines are fits to Eq. 2 ($x = 0.0017$) and a stretched exponential function for the remaining samples. The fit values for $\lambda_{ZF}$ are 8.3(1) µs$^{-1}$, 10.6(4) µs$^{-1}$, and 12.2(3) µs$^{-1}$ with $\beta$ values of 0.63(1), 0.73(3), and 0.82(2), for the $x = 0.0085$, 0.0406, and 0.0855 samples, respectively, and are consistent with values presented in Ref. 8. The $x = 0.0085$ sample shows behavior intermediate to Eq. 2 and a stretched exponential.

To determine the magnetic field necessary to completely suppress the FμF oscillations, we studied the longitudinal magnetic field dependence for the $x = 0.0017$ sample taken at low temperature and with the muon polarization parallel to the c-axis. Application of modest longitudinal (**B** // **c**) fields of 120 G is sufficient to suppress oscillations (Fig. 5). Also shown are the Quantum simulations, taken for $\Gamma = 8$ G, for applied longitudinal fields, which agree nicely with the data.

While our zero-field simulations show the muon response should cross over to a LKT response for higher concentrations and lower temperatures, only the sample with $x = 0.0085$ shows any sign of the characteristic minimum associated with LKT relaxation. This result is consistent with the conclusion presented in Ref. 8: for samples with Ho concentrations higher than $x = 0.01$, the fluctuations of the holmium magnetic moments remain strong at low temperatures, and this result justifies their use of the fluctuating LKT depolarization function to fit the data.

Based on our simulations of the high frequency AC susceptibility and μSR response for low concentration samples [14], we find that the Ho-Ho cross relaxation rate is of the same order of magnitude as, and has a comparable temperature dependence to, the fluctuation rate presented in Ref. 8. On the other hand, Ref. 14 also shows that cross relaxation becomes very slow in the limit T → 0 K, a result confirmed by the rapid suppression by magnetic field of the depolarization rate for $x = 0.0085$ at low temperatures [15]. These discrepancies point to the need for more sophisticated simulations to confirm that fluctuations are present in higher concentration samples, and if so, to reveal the nature of their driving mechanism.

As noted in Ref. 8, the Ho field distribution is anisotropic, and this will influence the detailed form of the depolarization curve. For a homogeneous distribution of paramagnetic ions in the low concentration limit, we can calculate the field distribution and anisotropy by using the expression for the magnetic resonance line half-width due to dipolar broadening [16]. In LiYF$_4$:Ho, a unit cell with volume $v_0$ contains two Y sites, and taking into account that $g_\perp = 0$ for the Ho$^{3+}$ ions ($g_{//} = 13.3$), we obtain

$$<B_z^2>^{1/2} = \frac{4}{3}<B_x^2>^{1/2} = \frac{8\pi^2 g_\| \mu_B x}{9\sqrt{3} v_0}. \tag{4}$$

In the case of a highly concentrated system,

$$\left\langle B_z^2 \right\rangle^{1/2} = \tfrac{1}{2} g_{//} \mu_B \left[ x \sum_k \left(1 - 3\frac{z_k^2}{r_k^2}\right)^2 r_k^{-6} \right]^{1/2}, \quad \left\langle B_x^2 \right\rangle^{1/2} = \tfrac{1}{2} g_{//} \mu_B \left[ x \sum_k \left(3\frac{x_k z_k}{r_k^2}\right)^2 r_k^{-6} \right]^{1/2} \quad (5)$$

where the sum is taken over Y sites at a distance $r_k$ from the muon, although here the contributions from the four nearest sites are omitted because the strong axial electric field induced by a muon splits the ground non-Kramers doublet of the holmium ion. The direct summation in Eq. 5 will slightly overestimate the field widths, since we are neglecting effects due to random crystal fields. The results of calculations in the low and high concentration limits are given in Table I, along with the field widths from Ref. 8 (and the one value estimated from this work). The ratio of the calculated width to the measured value increases from below 2 for x = 0.0017 up to nearly 7 for x = 0.12.

The calculated anisotropy of 4/3 is not very large, but nonetheless alters the results of fitting based on an isotropic distribution. Simulations were carried out for Lorentzian distributions in the isotropic and anisotropic cases, with $\left\langle B_x^2 \right\rangle = \left\langle B_y^2 \right\rangle = \tfrac{3}{4} \left\langle B_z^2 \right\rangle$ in the latter instance; $\left\langle B_x^2 \right\rangle + \left\langle B_y^2 \right\rangle + \left\langle B_z^2 \right\rangle$ was held constant for the two cases. For the muon polarization parallel (perpendicular) to the c-axis the depth of the characteristic minimum and the asymptotic limit of 1/3 are shifted upward (downward) by roughly 16% (8%), and we conclude that fitting the data with an isotropic fluctuating LKT provides only a qualitative measure of the field distribution. On the other hand, anisotropy effects cannot account for the very large discrepancy between the calculated distribution widths and those extracted from the curve fits. One interesting possibility is that the Ho-Ho correlations narrow the distribution widths.

The results presented here show that discerning the contribution to the depolarization due to the Ho ions from that of the FµF and more distant F nuclei can be complicated when the characteristic field distribution is less than 50 G, corresponding to $a = 4.1$ µs$^{-1}$. This is the case for our samples with $x = 0.0017$ and $0.0085$. Alternatively, an applied longitudinal field of more than roughly 120 G will decouple the muon from the fluorine nuclei and suppress FµF oscillations. These results allow us to confidently determine the region of concentration-magnetic field parameter space where we can ignore the effects of the fluorine-muon interaction, an important factor when studying low concentration samples for the onset of Ho magnetic correlations. Our LF relaxation measurements will be presented in a future publication, along with AC susceptibility measurements and detailed numerical simulations which include the effects of anisotropy, spin-phonon coupling, and Ho-Ho cross relaxation [14].

This work was supported by NSF grant DMR-0710525. Experiments were performed at the ISIS Rutherford Appleton Laboratories (UK) and the Paul Scherrer Institute (Switzerland). MJG would like to thank G. M. Luke and J. Rodriguez for helpful discussions.

Table I. Comparison of experimental and calculated values of the magnetic field distribution widths for Lorentzian ($a$) and Gaussian ($\Delta$) distributions, and the calculated distribution anisotropy. The experimental values are from Ref. 8, except for the $x = 0.0017$ value which is estimated in this work.

| $x$ | $a$ (μs$^{-1}$) | $\Delta$ (μs$^{-1}$) | $\langle B_z^2 \rangle^{1/2}$ (G) | $\langle B_z^2 \rangle^{1/2}$ (G) Eq. 4 | $\langle B_z^2 \rangle^{1/2}$ (G) Eq. 5 | $\langle B_x^2 \rangle / \langle B_z^2 \rangle$ |
|---|---|---|---|---|---|---|
| 0.0017 | 0.7 | -- | 4.7 | 7.4 | -- | 0.75 |
| 0.0180 | 4.5 | -- | 30.5 | 79 | -- | 0.75 |
| 0.0450 | 9.6 | -- | 65 | 197 | -- | 0.75 |
| 0.0800 | 12.6 | -- | 85 | 350 | -- | 0.75 |
| 0.12 | 11.8 | -- | 80 | 525 | 510 | 0.75 |
| 0.25 | -- | 17.7 | 208 | -- | 737 | 0.87 |

Figure captions

Figure 1. Temperature evolution of the depolarization curves for samples with (a) $x = 0.0017$ and (b) $x = 0.0854$. Solid lines are fits to Eqn. 2. In (b) $G_{F\mu F}(t) = 1$ for the curves at $T = 1.6$ K and 6 K. In (a) curve for T = 50 K would be nearly identical to T = 20 K and 14 K. Curves have been offset for clarity.

Figure 2. Temperature dependence of the fit parameters $\lambda_{ZF}$ (left axis), $\omega_D$ (right axis), and $\beta$ (inset) for the $x = 0.0017$ sample (muon polarization parallel to c-axis).

Figure 3. Simulated evolution of the FµF depolarization with increasing magnetic disorder $\Gamma$. Simulated data (symbols) are fitted with (solid lines) Eqn. 2 for 1 and 4 G, and Eqn. 3 for 16, 50 and 80 G.

Figure 4. Low temperature depolarization curves for samples with $x = 0.0017, 0.0085, 0.0406$ and $0.0854$ (muon polarization parallel to the c-axis). Solid lines are fits to Eqn. 2; $G_{F\mu F}(t) = 1$ for $x \geq 0.0085$. The curves have been offset for clarity.

Figure 5. Measured (open squares) evolution of the muon depolarization with increasing longitudinal field at low temperature and simulated (solid lines) FµF data over the same field range with magnetic disorder $\Gamma = 8$ G.

Figure 1

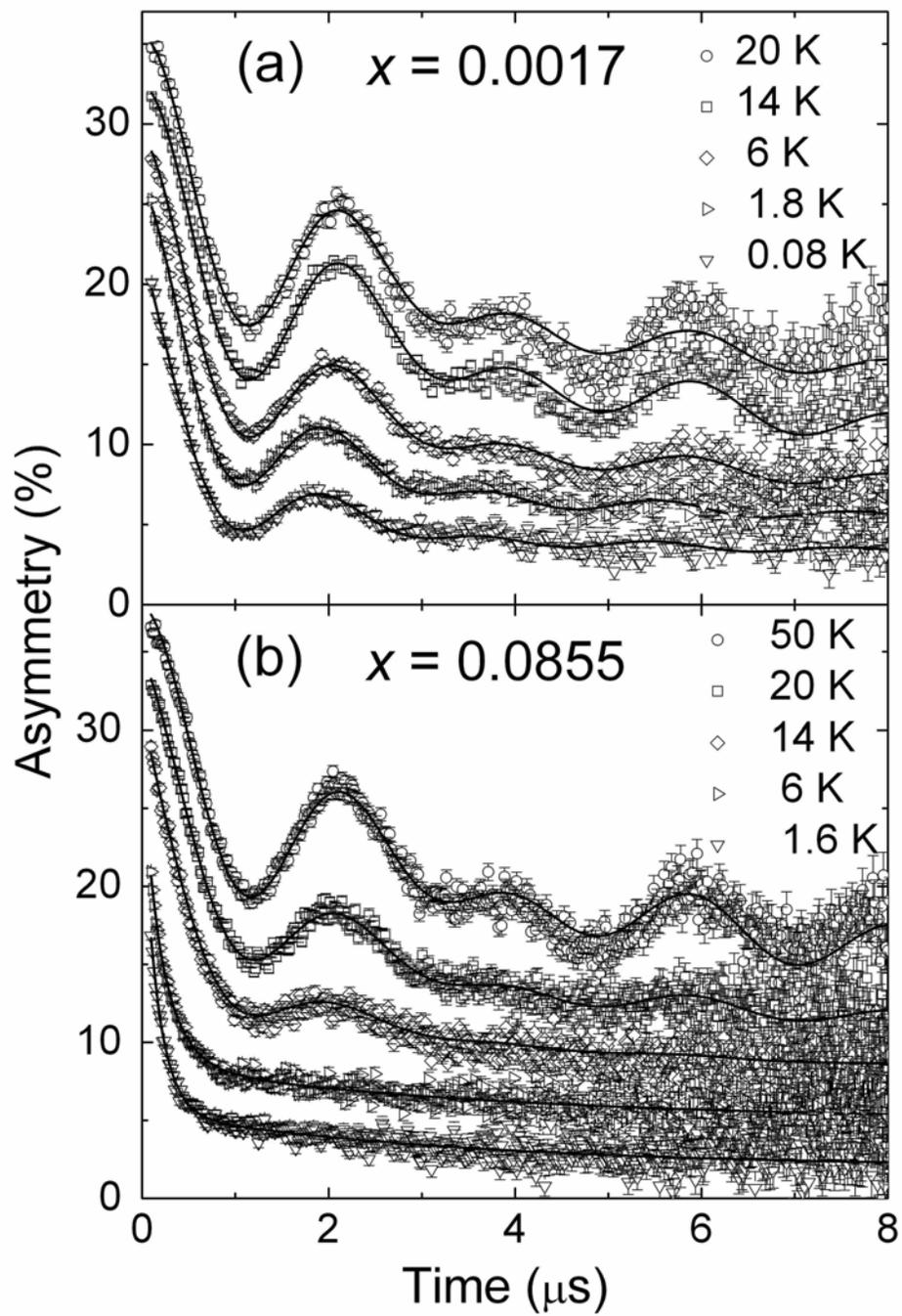

Figure 2

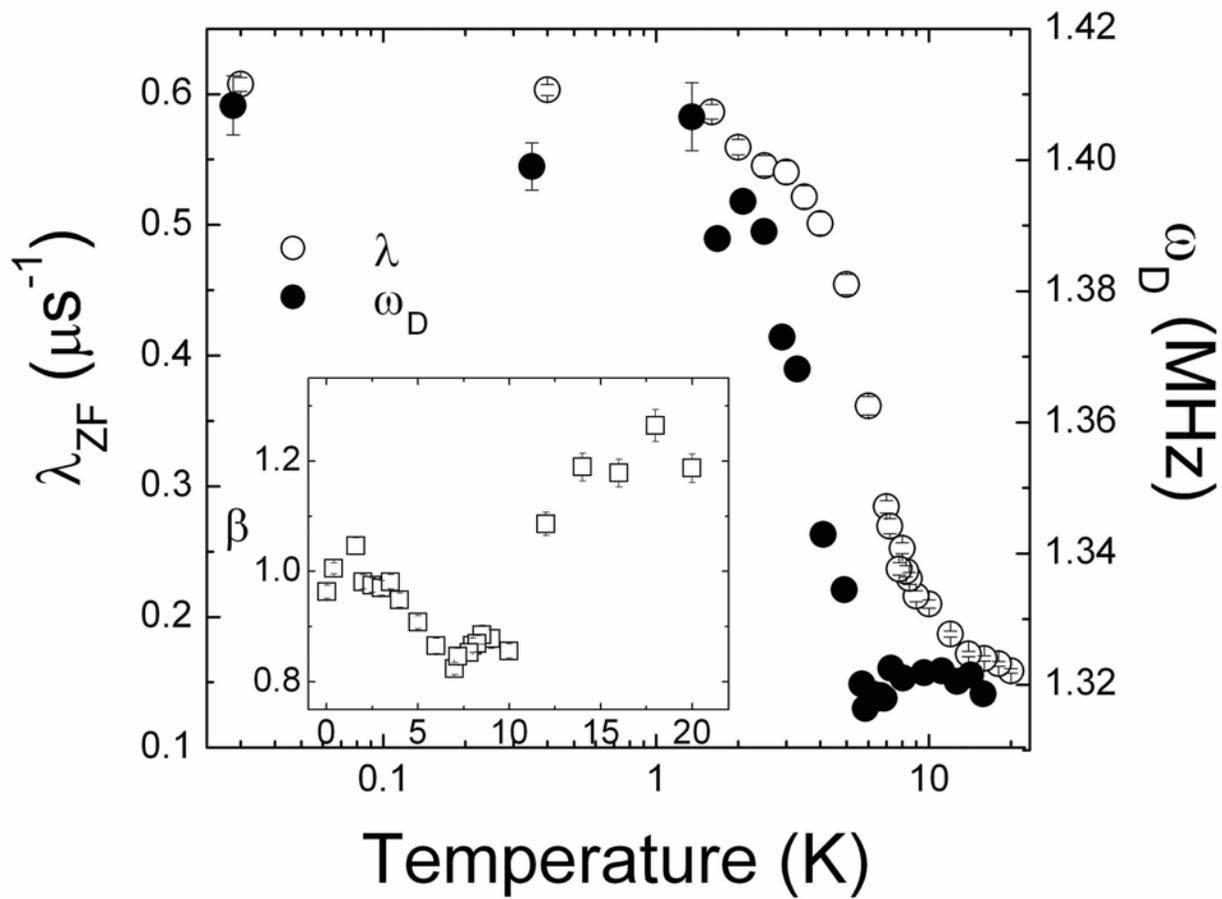

Figure 3

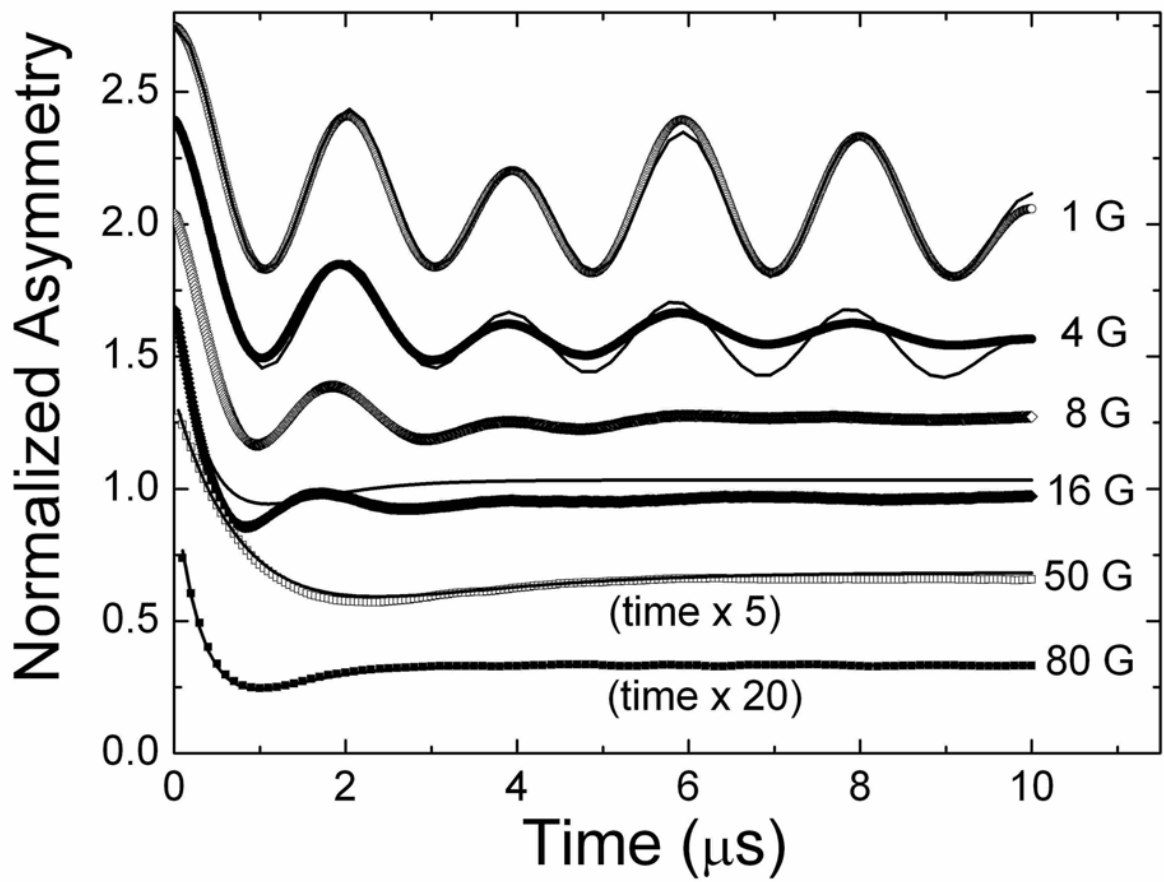

Figure 4

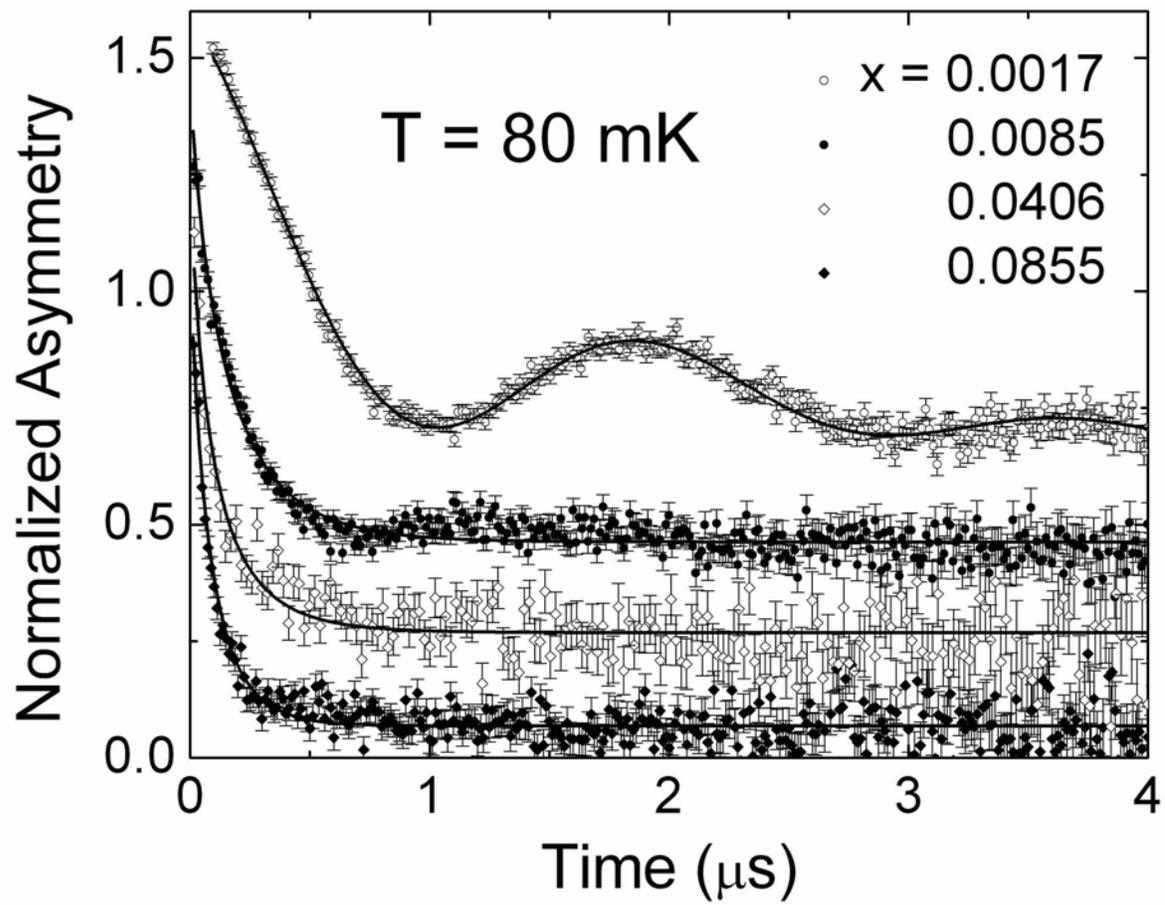

Figure 5

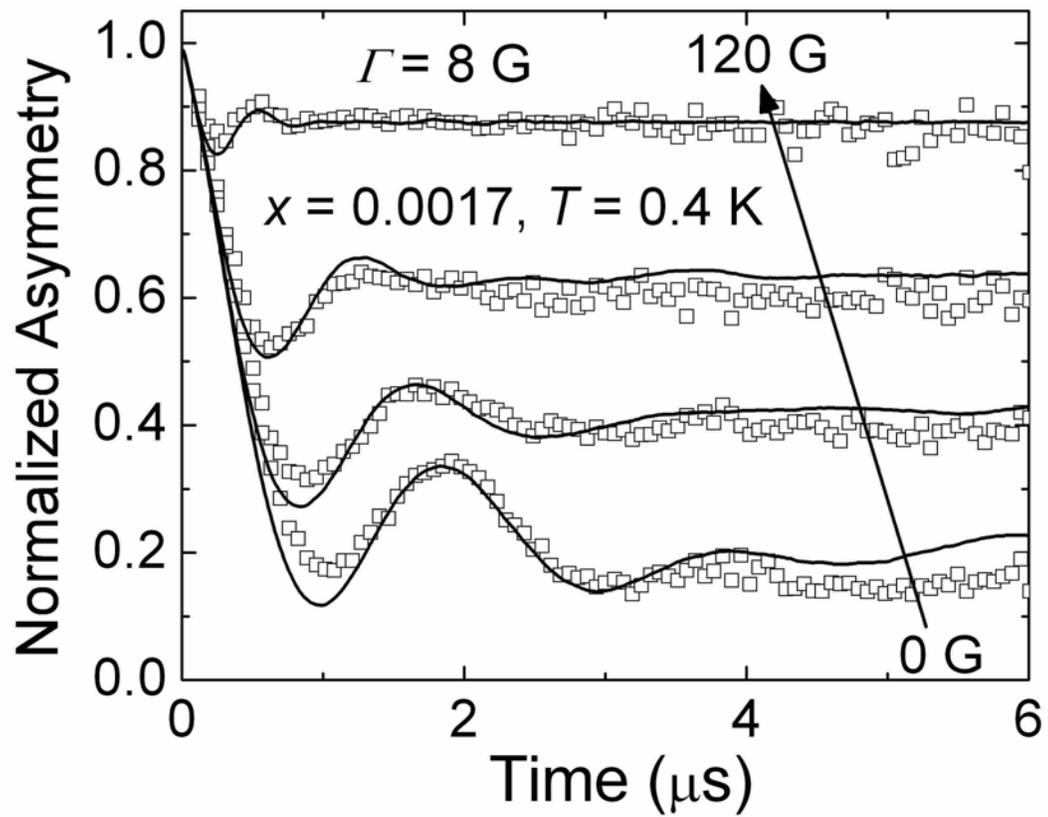